\documentclass[showkeys,12pt, preprint,preprintnumbers,nofootinbib, groupedaddress,superscriptaddress,amsmath,amssymb]{revtex4}
\usepackage{graphicx}% Include figure files
\usepackage{dcolumn}% Align table columns on decimal point
\usepackage{bm}% bold math
\usepackage{amssymb}
\usepackage{amsmath}
\usepackage{epsfig}    
\usepackage{color}
\usepackage{slashed}
\usepackage{hhline}
\usepackage{hyperref}
\usepackage{soul}
\def\be{\begin{equation}}
\def\ee{\end{equation}}
\newcommand{\bea}{\begin{eqnarray}}
\newcommand{\eea}{\end{eqnarray}}

\numberwithin{equation}{section}

\begin{document}

{\begin{flushright}{KIAS-P16086}
\end{flushright}}

%%%%%%%%%
\title{Dark matter physics in neutrino specific two Higgs doublet model  }
%\preprint{KIAS-P14078}
%
\author{Seungwon Baek}
\email{swbaek@kias.re.kr}
\affiliation{School of Physics, KIAS, Seoul 02455, Korea}
\author{Takaaki Nomura}
\email{nomura@kias.re.kr}
\affiliation{School of Physics, KIAS, Seoul 02455, Korea}

\date{\today}

\begin{abstract}
Although the seesaw mechanism is a natural explanation for the small neutrino masses, there are cases when the Majorana
mass terms for the right-handed neutrinos are not allowed due to symmetry. In that case, if neutrino-specific Higgs
doublet is introduced, neutrinos become Dirac particles and their small masses can be explained by its small VEV. 
We show that the same symmetry, which we assume a global $U(1)_X$, can also be used to explain the
stability of dark matter. In our model, a new singlet scalar breaks the global symmetry spontaneously down to a discrete
$Z_2$ symmetry. The dark matter particle, lightest $Z_2$-odd fermion, is stabilized. We discuss the phenomenology of dark
matter: relic density, direct detection, and indirect detection. We find that the relic density can be explained by a
novel Goldstone boson channel or by resonance channel. In the most region of parameter space considered, the direct
detections is suppressed well below the current
experimental bound. Our model can be further tested in indirect detection experiments such as FermiLAT gamma ray
searches or neutrinoless double beta decay experiments. 
\end{abstract}
\maketitle 
\newpage

\section{Introduction}
A natural scenario to explain the sub-eV neutrino masses is type-I seesaw mechanism in which very heavy standard model (SM)
singlet right-handed neutrinos are introduced. In this case the light-neutrinos become
Majorana particles and the scenario can be tested at neutrinoless double beta decay experiments.

A more straightforward way for the generation of neutrino masses in parallel with the generation of  quark or charged
lepton masses is just to introduce right-handed neutrinos to get Dirac neutrino masses with the assumption of lepton
number conservation to forbid the Majorana mass
terms of the right-handed neutrinos. The problem in this case is the neutrino Yukawa couplings should be tiny ($\lesssim
10^{-11}$) while the top quark Yukawa coupling is of order 1. To give Dirac masses to neutrinos, while avoiding this large hierarchy problem,
neutrino-two-Higgs-doublet model ($\nu$THDM) was suggested~\cite{Davidson:2009ha,Wang:2006jy}. 
In this model, the small neutrino masses are explained by the small VEV
of a second Higgs doublet ($v_1 = \sqrt{2} \langle \Phi_1^0 \rangle \sim {\cal O}(1) \,{\rm eV}$) while the neutrino Yukawa couplings can be of order 1.
The authors in Ref.~\cite{Davidson:2009ha} introduced global $U(1)$ symmetry, $U(1)_X$, which is softly broken to forbid 
Majorana mass terms of the right-handed neutrinos.
In their model, all the SM fermions except neutrinos obtain masses via Yukawa interactions with the SM-like Higgs doublet,
$\Phi_2$, while only neutrinos get masses from Yukawa interaction with $\Phi_1$:
\bea
{\cal L}_Y = -\overline{Q}_L Y^u \widetilde{\Phi}_2 u_R -\overline{Q}_L Y^d \Phi_2 d_R 
-\overline{L}_L Y^e \Phi_2 e_R -\overline{L}_L Y^\nu \widetilde{\Phi}_1 \nu_R +{h.c.},
\label{Orig_Y}
\eea
where $\widetilde{\Phi}_i = i \sigma_2 \Phi_i^* (i=1,2)$.
The $\Phi_1$ and $\nu_R$ are assigned with the global charge 1 under $U(1)_X$. The global symmetry forbids the Majorana mass term $
\nu_R \nu_R$. If the global symmetry is {\it softly} broken by introducing a term in the scalar potential, $ V \ni - m_{12}^2
\Phi_1^\dagger \Phi_2 + h.c.$, the small VEV is obtained by seesaw-like formulas
\bea
v_1 =\frac{m_{12}^2 v_2}{M_A^2},
\eea
where $M_A$ is the pseudo-scalar mass~\cite{Davidson:2009ha}. For the electroweak scale $M_A (\sim 100)$ GeV, $v_1 \sim
1$ eV can be obtained by $m_{12} \sim {\cal O}(100)$ keV.

We extend the model to include a natural dark matter  (DM) candidate, $\psi$. In our model the global symmetry, $U(1)_X$, is {\it spontaneously} broken
down to discrete $Z_2$ symmetry by VEV of a new singlet scalar, $S$. The remnant $Z_2$ symmetry makes the dark matter
candidate stable. The resulting Goldstone boson provides a new annihilation channel for the DM relic density.
It is feebly coupled to the SM particles due to tiny $v_1$, avoiding experimental constraints.
We also study the DM direct detection and indirect detection. They are typically well below the current experimental sensitivity.

The paper is organized as follows. In Section~\ref{sec:model}, we introduce our model.
In Section~\ref{sec:DM}, we study DM phenomenology in our model: relic abundance, direct and indirect detection
of the DM. In Section~\ref{sec:con}, we conclude.

%%%%%%%%%%%%%%%%%%%%%%%%%%%%%%%%%%%%%
\section{The Model}
\label{sec:model}

\begin{widetext}
\begin{center} 
\begin{table}%[tbc]
%\begin{tiny}
\begin{tabular}{|c||c|c|c||c|c|}\hline\hline  
&\multicolumn{3}{c||}{Scalar Fields} & \multicolumn{2}{c|}{New Fermion} \\\hline
& ~$\Phi_1$~ & ~$\Phi_2$~ & ~$S$ ~ & ~$\nu_R$  & ~$\psi$~  \\\hline 
$SU(2)_L$ & $\bm{2}$  & $\bm{2}$  & $\bm{1}$ & $\bm{1}$ & $\bm{1}$  \\\hline 
$U(1)_Y$ & $\frac12$ & $\frac12$  & $0$ & $0$ & $0$   \\\hline
 $U(1)_X$ & $2$ & $0$   & $2$ & $2$ & $1$     \\\hline
%%%
\end{tabular}
\caption{Scalar fields and new fermion in our model where $\nu_R$ is Majorana type while $\psi$ is Dirac type.}
\label{tab:1}
% \end{tiny}
\end{table}
\end{center}
\end{widetext}

In this section, we introduce our model which is an extension of the model given in Ref.~\cite{Davidson:2009ha}. The
scalar field contents and new fermions are summarized in Table.~\ref{tab:1} where we also show the charge assignments under global $U(1)_X$
symmetry.  
We can write $U(1)_X$-invariant as well as the SM-gauge invariant scalar potential, Yukawa interactions for the leptons and
new fields as
\begin{align}
\label{eq:potential}
V(\Phi_1, \Phi_2, S) = & - m_{11}^2 \Phi^\dagger_1 \Phi_1 - m_{22}^2 \Phi_2^\dagger \Phi_2 - m_{S}^2 S^\dagger S - (\mu \Phi^\dagger_1 \Phi_2 S + h.c.) \nonumber \\
& + \lambda_1 (\Phi_1^\dagger \Phi_1)^2 + \lambda_2 (\Phi_2^\dagger \Phi_2)^2 + \lambda_3 \Phi_1^\dagger \Phi_1 \Phi_2^\dagger \Phi_2 
+ \lambda_4 \Phi_1^\dagger \Phi_2 \Phi_2^\dagger \Phi_1 + \lambda_S (S^\dagger S)^2 \nonumber \\
&+ \lambda_{1S} \Phi_1^\dagger \Phi_1 S^\dagger S + \lambda_{2S} \Phi_2^\dagger \Phi_2 S^\dagger S, \\
\label{eq:Yukawa}
{\cal L}  \supset & - y_{ij}^e \bar L_i \Phi_2 e_{Rj} - y^\nu_{ij} \bar L_i  \tilde \Phi_1 \nu_{Rj} + h.c, \\
\label{eq:Lpsi}
{\cal L} \supset & \, \bar \psi i \gamma^\mu \partial_\mu \psi - m_\psi \bar \psi \psi - \frac{f}{2} \bar \psi^c \psi S^\dagger - \frac{f^*}{2} \bar \psi \psi^c S.
\end{align}
%where $\tilde \Phi_1 = i \sigma_2 \Phi_1^*$. 
Thus Dirac masses of neutrinos are generated by VEV of $\Phi_1$ which is assumed to be much smaller than electroweak scale to obtain tiny neutrino mass~\cite{Davidson:2009ha, Wang:2006jy}.
In addition, a $Z_2$ symmetry remains when $U(1)_X$ is broken by non-zero VEVs of $S$. Note that only $\psi$ is $Z_2$ odd particle while other particles including those in SM sector are even under the $Z_2$, guaranteeing stability of $\psi$. Thus $\psi$ can be a DM candidate in the model. 

{Here we note that a global symmetry is considered to be broken by quantum effect at Planck scale, $M_{\rm pl}$.  In such
  a case we would have a Planck suppressed effective operator $\bar \psi \tilde H^\dagger (\slashed{D} L)$ which breaks
  the $Z_2$ symmetry from global $U(1)_X$ in our model inducing instability of DM candidate~\cite{Mambrini:2015sia}. 
Then the lifetime of $\psi$ becomes too short to be dominant component of DM for $m_\psi \gtrsim O(0.1)$ GeV if the
global $U(1)_X$ breaking operator is not suppressed by very small dimensionless coupling.
This instability could be avoided assuming our global $U(1)_X$ is a subgroup of some gauge symmetry broken at scale higher
than electroweak scale. Another way is introducing local $U(1)_{B-L}$ symmetry where $\psi$ does not have $B-L$ charge,
in order to forbid the operator inducing decay of $\psi$. In this paper, we just assume our DM candidate is stabilized
by the $Z_2$ from $U(1)_X$. On the other hand, the breaking of the global $U(1)_X$ at Planck scale does not affect
neutrino mass since such a contribution is highly suppressed by the factor of $v/M_{pl} \sim 10^{-16}$.}

The scalar fields can be written by
\begin{equation}
 \Phi_1 = \begin{pmatrix} \phi^+_1 \\ \frac{1}{\sqrt{2}} (v_1 + h_1 + i a_1) \end{pmatrix}, \quad
 \Phi_2 = \begin{pmatrix} \phi^+_2 \\ \frac{1}{\sqrt{2}} (v_2 + h_2 + i a_2) \end{pmatrix}, \quad S = \frac{1}{\sqrt{2}} r_S e^{ i\frac{a_S}{v_S}}.
\end{equation}
Note that we write $S$ in terms of radial field $r_S = v_S + \rho$ and phase field $a_S$ with $\langle a_S \rangle =0$~\cite{Weinberg:2013kea} since $a_S$ becomes physical Goldstone boson as shown below.
The VEVs of the scalar fields are obtained by requiring $\partial V(v_1,v_2,v_S)/\partial v_i =0$ which provides following conditions:
\begin{align}
\label{eq:v1}
& -2 m_{11}^2 v_1 + 2 \lambda_1 v_1^3 + v_1 (\lambda_{1S} v_S^2 + \lambda_3 v_2^2 + \lambda_4 v_2^2) - \sqrt{2} \mu v_2  v_S =0, \\
& -2 m_{22}^2 v_2 + 2 \lambda_2 v_2^3 + v_2 (\lambda_{2S} v_S^2 + \lambda_3 v_1^2 + \lambda_4 v_1^2) - \sqrt{2} \mu v_1 v_S =0, \\
& -2 m_{SS}^2 v_S + 2 \lambda_S v_S^3 + v_S (\lambda_{1S} v_1^2 + \lambda_{2S} v_2^2 ) - \sqrt{2} \mu v_1 v_2 =0.
\end{align}
We then find that these conditions can be satisfied with $v_1 \simeq \mu \ll \{ v_2, v_S \}$ and SM Higgs VEV is given
as $v \simeq v_2 \simeq 246$ GeV. From (\ref{eq:v1}) we find that $v_1$ is proportonial to and of the same order with $\mu$:
\bea
v_1 \simeq \frac{\sqrt{2} \mu v_2 v_S}{\lambda_{1S} v_S^2 +(\lambda_3+\lambda_4) v_2^2 -2 m_{11}^2}.
\eea
{Typically $v_1 \sim \mu$ is required for electroweak scale $v_2, v_S$. Taking neutrino
  mass scale as $m_\nu \sim 0.1$ eV, the value of $\mu/v_2$ should be $\mu/v_2 \sim {\mathcal O}(10^{-12})[{\mathcal
    O}(10^{-6})]$ when the order of the Yukawa coupling $Y^\nu$ is ${\mathcal O}(1)[ {\mathcal O}(10^{-6}) (\sim
  m_e/v_2)]$. }
We note, however, that small $\mu (\ll v)$ is technically natural~\cite{tHooft,Baek:2014sda} because $\mu \equiv 0$ enhances the symmetry of the Lagrangian
(\ref{eq:potential}) to additional $U(1)$ under which only the $S$ field is charged while all the others are neutral.

Here we consider masses and mass eigenstate of the scalar sector by analyzing the scalar potential with $v_1 \sim \mu \ll \{ v_2, v_S \}$. \\
{\it Pseudo-scalar} : 
Mass matrix for pseudo-scalars is given, in the basis of $(a_1, a_2, a_S)$, by
\begin{equation}
M^2_A \simeq \frac{\mu}{\sqrt{2}} \begin{pmatrix} \frac{v_2 v_S}{v_1} & - v_S & - v_2 \\ -v_S & \frac{v_1 v_S}{v_2} & v_1 \\ -v_2 & v_1 & 0 \end{pmatrix} 
 \simeq  \begin{pmatrix} \frac{\mu v_2 v_S}{\sqrt{2} v_1} & 0 & 0 \\ 0 & 0 & 0 \\ 0 & 0 & 0 \end{pmatrix},
\end{equation}
where we used $S \simeq (v_S + \rho + i a_S)/\sqrt{2}$ to obtain the mass matrix.
We thus find three mass eigenstates $A$, $a$, and $G_0$:  $A (\simeq a_1)$ is massive pseudo-scalar, 
$a (\simeq a_S)$ is physical massless Goldstone boson associated with $U(1)_X$ breaking as indicated above, 
and $G_0 (\simeq a_2)$ is massless Nambu-Goldstone (NG) boson which is absorbed by $Z$ boson.
The mass of $A$ is given by
\bea
m_A^2 =\frac{\mu(v_1^2 v_2^2 + v_1^2 v_S^2 + v_2^2 v_S^2)}{\sqrt{2} v_1 v_2 v_S}  \simeq 
\frac{\mu v_2 v_S}{\sqrt{2} v_1}.
\eea
Note that the existence of physical Goldstone boson $a$ does not lead to serious problems in particle physics or
cosmology since it does not couple to SM particles directly except to SM Higgs.
Invisible decay width of $Z$-boson strongly constrains the $Z \to H_i a$ decay\footnote{$H_i (i=1,2,3)$ are neutral scalars
defined below.}.
Since $v_S$ is a free parameter, we can make $\rho$ (or the mass eigenstate with $\rho$ as main component)
heavier than the $Z$-boson mass to evade the problem~\cite{Frampton:2002rn}.
In our model, $a$ can couple also to electron via $ i g_{\bar{e} e a} \, a \, \bar{e} \gamma_5 e$ interaction through mixing with
the SM Higgs doublet.
Stellar energy loss constrains $g_{\bar{e} e a} \lesssim 10^{-12}$ model-independently~\cite{Chang:1988aa}.
The tree-level contribution in our model, $g_{\bar{e} e a} \simeq m_e v_1/(v v_S) \approx 2 \times 10^{-16} (v_1/ 1 \,
{\rm eV}) (100 \, {\rm GeV}/v_S)$, satisfies the bound safely.

%Also contribution to the effective number of neutrino species $\Delta N_{\rm eff}$ would be small since $a$ decouples in
%early Universe.  
Our model can also contribute about 0.39 to the effective number of neutrino species 
$\Delta N_{\rm eff}$~\cite{Weinberg:2013kea} when $\lambda_{2S}=0.005$ and $m_{H_3}=500$ MeV.
This can solve~\cite{Riess:2016jrr,Ko:2016uft} about 3.4$\sigma$ discrepancy between Hubble Space Telescope~\cite{Riess:2016jrr} 
and Plank~\cite{Ade:2015xua} in the measurement of Hubble constant.  
Since the mechanism is almost the same with that detailed in~\cite{Weinberg:2013kea} we do not further discuss implication of the
Goldstone boson on $\Delta N_{\rm eff}$.
\\
{\it Charged scalar} : 
For charged scalar case, mass matrix in the basis of $(\phi_1^\pm, \phi_2^\pm)$ is given by
\begin{equation}
M^2_{H^\pm} = \begin{pmatrix} \frac{v_2 (\sqrt{2} \mu v_S - \lambda_4 v_1 v_2 )}{2v_1} & - \frac{1}{2} (\sqrt{2}\mu v_S - \lambda_4 v_1 v_2)  \\ 
- \frac{1}{2} (\sqrt{2}\mu v_S - \lambda_4 v_1 v_2) & \frac{v_1 (\sqrt{2} \mu v_S - \lambda_4 v_1 v_2)}{2 v_2}  \end{pmatrix} 
 \simeq  \begin{pmatrix} \frac{v_2 (\sqrt{2} \mu v_S - \lambda_4 v_1 v_2 )}{2v_1}& 0 &  \\ 0 & 0 &  \end{pmatrix},
\end{equation}
which indicates that $\phi_1^\pm$ is approximately physical charged scalar, $H^\pm$, and $\phi^\pm_2$ is approximately $G^\pm$,  the
NG boson absorbed by $W^\pm$ boson. We obtain the charged Higgs mass as
\bea
m_{H^\pm}^2 = \frac{(v_1^2+v_2^2)(\sqrt{2} \mu v_S -\lambda_4 v_1 v_2)}{2 v_1 v_2} \simeq
\frac{v_2 (\sqrt{2} \mu v_S - \lambda_4 v_1 v_2 )}{2v_1}.
\eea
\\
{\it CP-even scalar} :
In the case of CP-even scalar, all three components are physical, and the mass matrix in the basis of $(h_1,h_2, \rho)$ is written as
\begin{align}
M^2_H &= \begin{pmatrix} 2 \lambda_1 v_1^2 + \frac{\mu v_2 v_S}{\sqrt{2} v_1} & (\lambda_3 + \lambda_4) v_1 v_2 - \frac{\mu v_S}{\sqrt{2}} & \lambda_{1S} v_1 v_S - \frac{\mu v_2}{\sqrt{2}} \\ 
(\lambda_3 + \lambda_4) v_1 v_2 - \frac{\mu v_S}{\sqrt{2}} & 2 \lambda_2 v_2^2 + \frac{\mu v_1 v_S}{\sqrt{2} v_2} & \lambda_{2S} v_2 v_S - \frac{\mu v_1}{\sqrt{2}} \\ 
\lambda_{1S} v_1 v_S - \frac{\mu v_2}{\sqrt{2}}  & \lambda_{2S} v_2 v_S - \frac{\mu v_1}{\sqrt{2}} & 2 \lambda_S v_S^2 + \frac{\mu v_1 v_2}{\sqrt{2} v_S} \end{pmatrix}  \nonumber \\
& \simeq \begin{pmatrix} \frac{ \mu v_2 v_S }{\sqrt{2} v_1} & 0 & 0 \\ 0 & 2 \lambda_2 v_2^2 & \lambda_{2S} v_2 v_S \\ 0 & \lambda_{2S} v_2 v_S & 2 \lambda_{S} v_S^2 \end{pmatrix}.
\end{align}
We find that all the masses of the mass eigenstates, $H_i (i=1,2,3)$, are at the electroweak scale and the mixings
between $h_1$ and other components are negligibly small while
the $h_2$ and $\rho$ can have sizable mixing. The mass eigenvalue and mixing angle for $h_2$ and $\rho$ system are given by
\begin{align}
& m_{H_2,H_3}^2 = \frac{1}{2} \left[ m_{22}^2 + m_{33}^2 \mp \sqrt{(m_{22}^2-m_{33}^2)^2 + 4 m_{23}^4} \right], \\
& \tan 2 \theta = \frac{-2 m_{23}^2}{m_{22}^2 - m_{33}^2}, \\
& m_{22}^2 = 2 \lambda_2 v_2^2, \quad m_{33}^2 = 2 \lambda_{S} v_S^2, \quad m_{23}^2 = \lambda_{2S} v_2 v_S.
\end{align}
Then mass eigenstates are obtained as 
\begin{equation}
\begin{pmatrix} H_1 \\ H_2 \\ H_3 \end{pmatrix} \simeq \begin{pmatrix} 1 & 0 & 0 \\ 0 & \cos \theta & - \sin \theta \\ 0 & \sin \theta & \cos \theta \end{pmatrix} \begin{pmatrix} h_1 \\ h_2 \\ \rho 
\label{eq:eigenstates}
\end{pmatrix}
\end{equation}
Note that $H_2$ is the SM-like Higgs, $h$, and $m_{H_2} \simeq m_{h}$ where mixing angle $\theta$ is constrained to be $\sin \theta \lesssim 0.2$ by data of Higgs search at the LHC~\cite{hdecay, Chpoi:2013wga, Cheung:2015dta,Cheung:2015cug}.
For small mixing, we have $H_2 \simeq h$ and $H_3  \simeq \rho$.
Note also that our case realizes alignment limit $\beta - \alpha \simeq \pi/2$ in two Higgs doublet sector which is consistent with current SM Higgs analysis~\cite{Benbrik:2015evd}.
In addition, we take into account constraint from $h \to a a$ decay which is induced by interaction term $1/(v_S) \rho \partial_\mu a \partial^\mu a$ from kinetic term of $S$.
The decay width can be estimated as 
\begin{equation}
\Gamma_{h \to a a} = \frac{\sin^2 \theta}{16 \pi} \left( \frac{m_h}{v_S} \right)^2 m_h,
\end{equation}
and we require upper limit of the branching ratio as $BR(h \to aa) < 0.23$ based on constraint of invisible decay of SM Higgs~\cite{Aad:2015txa,Aad:2015pla, Khachatryan:2016whc}. 
The phenomenology of two Higgs doublet sector and constraints are discussed in Ref.~\cite{Davidson:2009ha, Machado:2015sha,Bertuzzo:2015ada} in detail. We thus focus on DM physics in the following analysis.

{\it Dark sector} : 
To obtain interactions of $\psi$ and physical scalar bosons, we define a field $\psi'$ by~\cite{Weinberg:2013kea} 
\begin{equation}
 \psi = \psi' e^{i \frac{a_S}{2 v_S}},
\end{equation}
so that the direct coupling of $a_S$ to $\psi'$ disappears.
Then the Lagrangian for $\psi'$ becomes 
\begin{equation}
\label{eq:Lpsi2}
{\cal L} \supset  \bar \psi' i \gamma^\mu \partial_\mu \psi' - m_\psi \bar \psi' \psi' - \frac{1}{2v_S} \bar \psi'
\gamma^\mu \psi' \partial_\mu a_S  - \frac{f}{2\sqrt{2}} \bar \psi^{'c} \psi' r_S - \frac{f}{2\sqrt{2}} \bar \psi'
\psi'^{c} r_S,
\end{equation}
where $f$ is taken to be real and positive by an appropriate choice of phase of $\psi$.
Since $r_S (= v_S + \rho)$ has non-zero VEV, the mass eigenstates of $Z_2$ odd fermions are obtained as a pair of self-charge-conjugate fields;
\begin{equation}
 \psi_+ = \frac{1}{\sqrt{2}} \left( \psi' +\psi'^{c} \right),  \quad \psi_- = \frac{-i}{\sqrt{2}} \left( \psi' -\psi'^{c} \right),
\end{equation}
which satisfy Majorana conditions $\psi_\pm^c = \psi_\pm$ and have mass eigenvalues 
\begin{equation}
m_\pm = m_\psi \pm \frac{f v_S}{\sqrt{2}}.
\end{equation}
Thus $\psi_-$ is our DM candidate in the following analysis.
Finally the Lagrangian for the mass eigenstates is given by
\begin{align}
{\cal L} \supset & \frac{1}{2} \sum_{\alpha = \pm} \bar \psi_\alpha \left[ i \gamma^\mu \partial_\mu  - m_\pm \right] \psi_\alpha 
-  \frac{i}{4 v_S} \left[ \bar \psi_+ \gamma^\mu \psi_- - \bar \psi_- \gamma^\mu \psi_+ \right] \partial_\mu a_S \nonumber \\
& - \frac{f}{2 \sqrt{2}} \rho \left[ \bar \psi_+ \psi_+ - \bar \psi_-  \psi_- \right]. 
\end{align}

%%%%%%%%%%%%%%%%%%%%%%%%%%%%%%%%%%%%%%%%%%%%%%%%%%%%%%%%%%%%%%%%%%%%%%%%%%%
\section{Dark matter physics}
\label{sec:DM}

\begin{figure}[t] 
\begin{center}
\includegraphics[width=40mm]{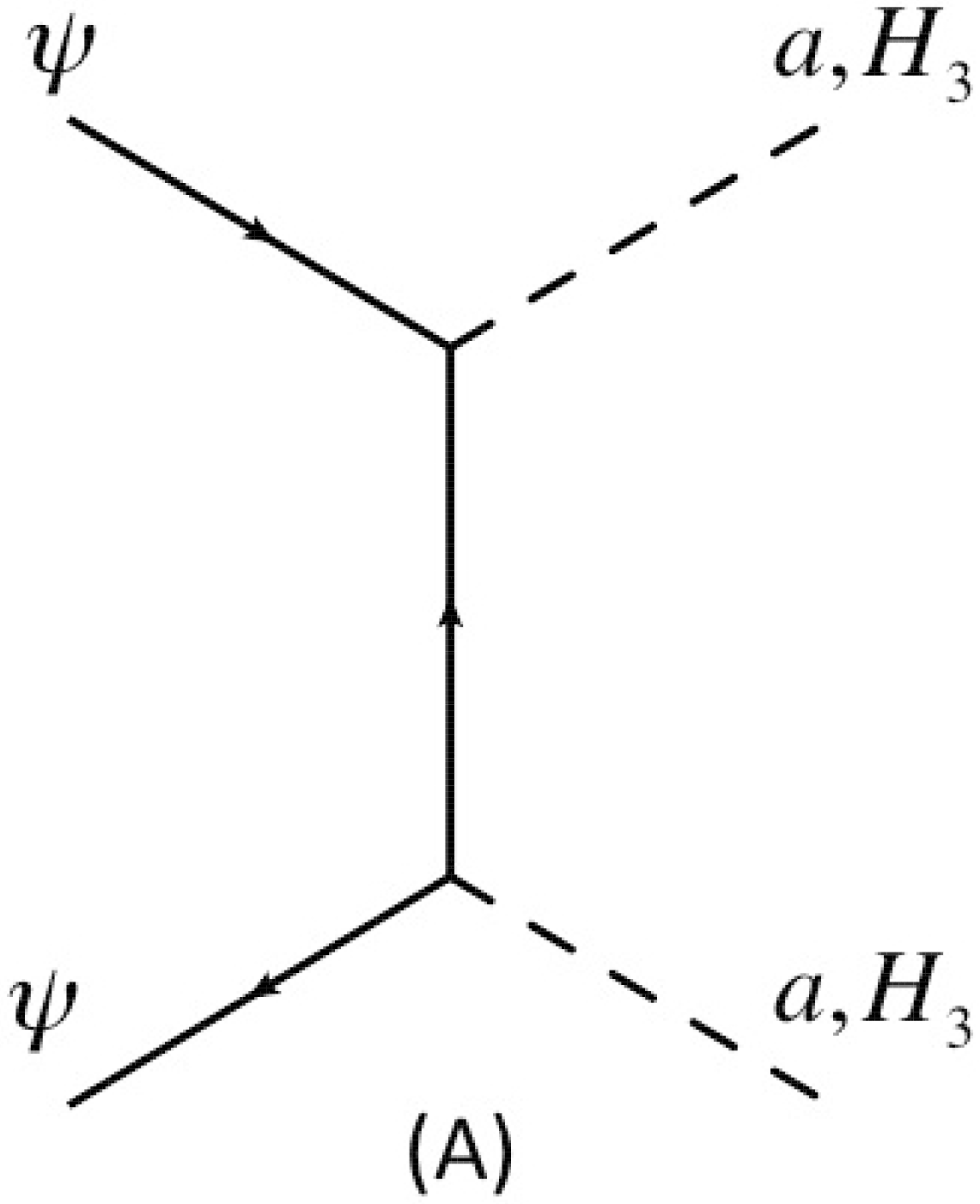} 
\includegraphics[width=60mm]{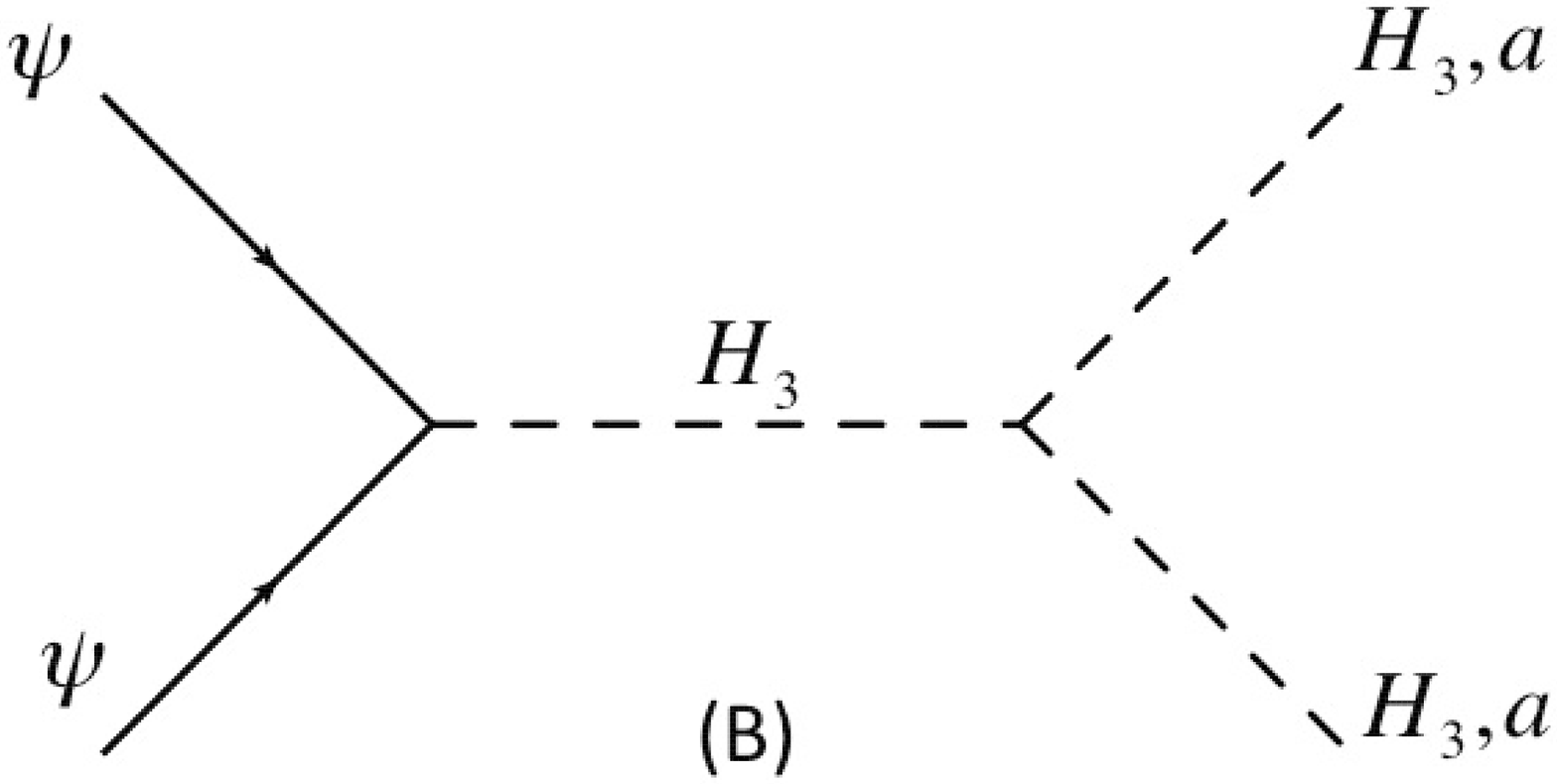} 
\includegraphics[width=60mm]{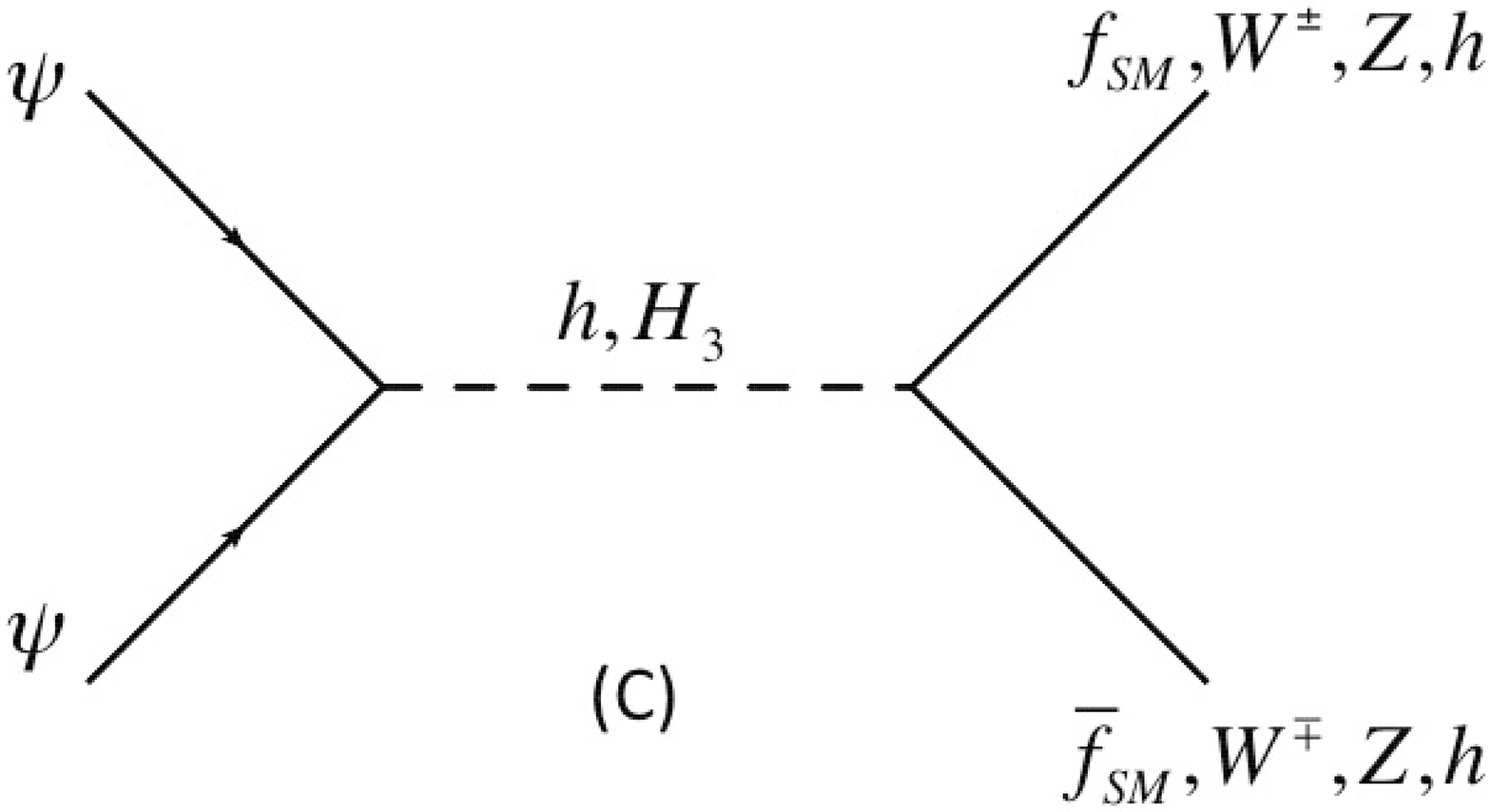}
\includegraphics[width=60mm]{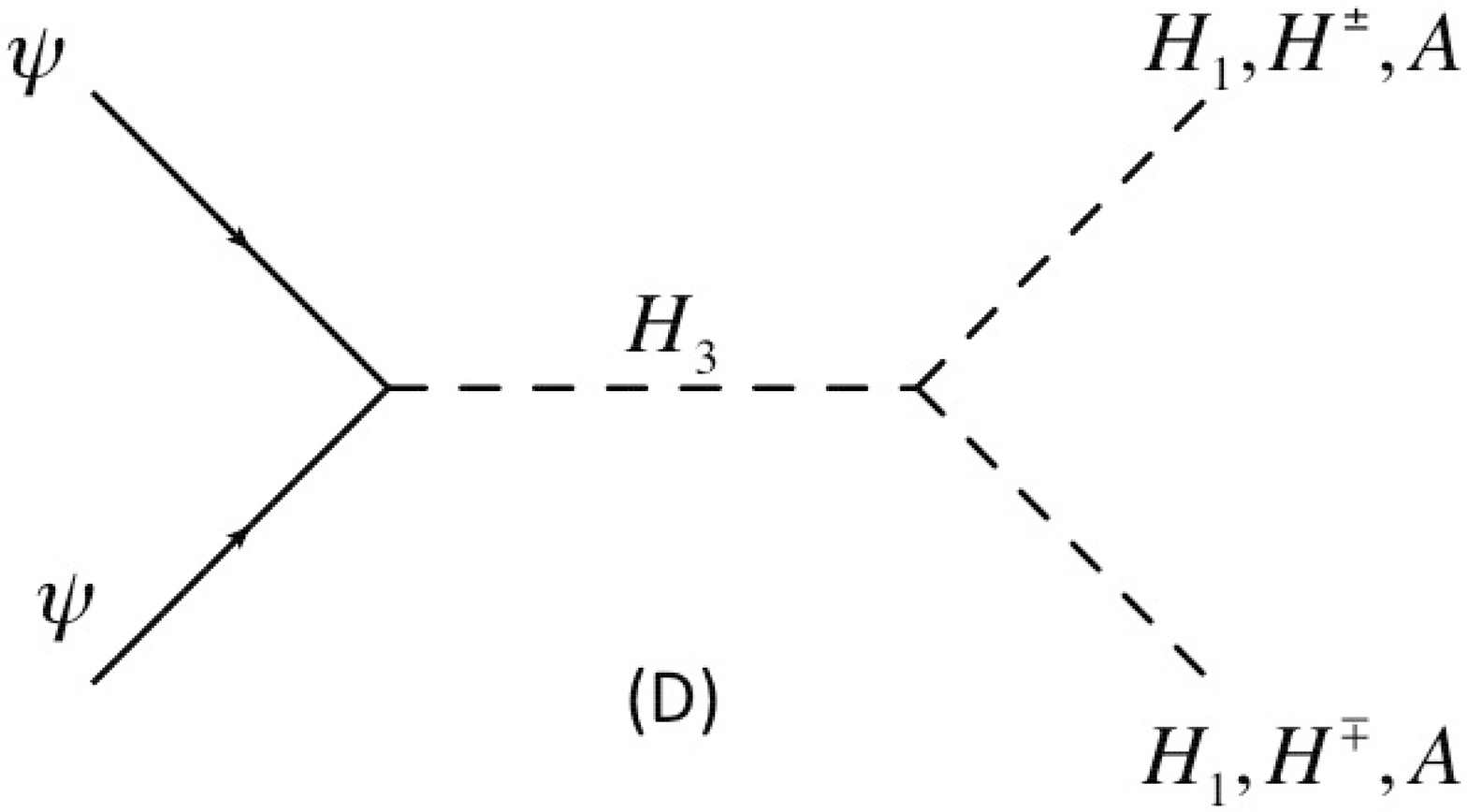}  
\end{center}
 
\caption{ The DM annihilation processes. 
\label{fig:DM}}
\end{figure}

In this section, we discuss DM physics such as relic density, direct detection and indirect detection.
Our DM candidate is the new Majorana fermion $\psi_-$ which is stable due to $Z_2$ symmetry as a remnant of the global $U(1)_X$ symmetry.
Interactions relevant to DM physics are obtained from the kinetic term of $S$,  terms in Eq.~(\ref{eq:potential}), and (\ref{eq:Lpsi2}):
\begin{align}
{\cal L} \supset & -\frac{f}{2\sqrt{2}} \rho (\bar \psi_+ \psi_+ - \bar \psi_- \psi_-) 
- \frac{i}{4 v_S } \left[ \bar \psi_+ \gamma^\mu \psi_- - \bar \psi_- \gamma^\mu \psi_+ \right] \partial_\mu a \nonumber \\
& - \mu_{SS} \rho^3 + \frac{1}{v_S} \rho \partial_\mu  a \partial^\mu a  - \mu_{1S} \rho \left(\phi^+_1 \phi^-_1 + \frac{1}{2} (h_1^2 + a_1^2) \right) - \frac{\mu_{2S}}{2} \rho h_2^2,
\end{align}
where we defined $\mu_{SS} \equiv \lambda_S v_S$, $\mu_{1S} \equiv \lambda_{1S} v_S$ and $\mu_{2S} \equiv \lambda_{2S} v_S$, and $\rho(h_2)$ can be written in terms of mass eigenstates via Eq.~(\ref{eq:eigenstates}).
In the following analysis, we consider four different scenarios for the coupling constants: (I) $f \leq \sqrt{4 \pi}$ and $\mu_{1S, 2S, SS} \ll 0.1 $ GeV,
(II) $f  \leq \sqrt{4 \pi}$ and $\mu_{SS} \gg \mu_{1S, 2S}$, (III) $f \leq 0.8$ and $\mu_{2S} \gg \mu_{1S, SS}$, (IV) $f  \leq 0.8$ and
$\mu_{1S} \gg \mu_{2S, SS}$.
For scenario (I), DM dominantly annihilate into $\rho \rho$ and/or $aa$ via interaction with coupling $f$ as
Fig.~\ref{fig:DM}-(A)~\cite{Lindner:2011it,Baek:2012ub,Weinberg:2013kea,Baek:2013ywa}  and $aa$ via process in Fig.~\ref{fig:DM}-(B).  
In the scenario (II), final states of DM annihilation process is same as scenario (I) where $\psi_- \psi_- \to H_3 \to H_3 H_3$ mode in Fig.~\ref{fig:DM}-(B) is added.
In the scenarios (III) and (IV), a DM pair dominantly
annihilates via s-channel processes where
$\rho (\simeq H_3)$ propagates as an intermediate particle; the dominant final states are, depending on parameters,
$\{h h, f_{SM}f_{SM}, W^+ W^-, ZZ \}$ and $\{H_1 H_1, A A, H^+ H^- \}$ for the scenarios (III) and (IV) respectively as shown in
Fig.~\ref{fig:DM} (C) and (D), and $a a$ channel in Fig.~\ref{fig:DM}-(B) which contributes to both scenarios.  
Note that, $\mu_{2S}$ induces mixing between $h_2$ and $\rho$ and we discuss constraint from direct detection taking into
account Higgs portal interaction \cite{Kim:2008pp,Baek:2011aa,Baek:2012se,Baek:2013qwa,Baek:2014jga,Chen:2015nea,Baek:2016kud} with the mixing
effect for scenario (III).
%%%%%%%%%%%%%%%%%%%%%%%%%%%%%%%%%%%%%%%%%%%%%%%%%%%%%%%%%
%%%%%%%%%%%%%%%%%%%%%%%%%%%%%%%%%%%%%%%%%%%%%%%%%%%%%%%%%%

\subsection{Relic density}
We estimate the thermal relic density of DM numerically using {\tt micrOMEGAs 4.3.1 }~\cite{Belanger:2014vza} to solve
the Boltzmann equation by implementing relevant 
interactions inducing the DM pair annihilation processes. Then we search for parameter sets which satisfy the
approximate region for the relic density~\cite{Ade:2013zuv}
\begin{equation}
0.11 \lesssim \Omega h^2 \lesssim 0.13.
\end{equation}
In numerical calculations random parameter sets are prepared in the following parameter ranges for each scenario:
\begin{align}
 {\rm For \, all \, scenario:} & \quad m_- \in [50, 1100] \ {\rm GeV}, \quad m_{H_3} \in [30, 2200], \quad v_S = 1000 \ {\rm GeV}, \\
 {\rm scenario \, (I):} & \quad f \in [0.1, \sqrt{4 \pi}], \quad  \mu_{1S}= \mu_{2S}= \mu_{SS} = 10^{-3} \ {\rm GeV},  \\
 {\rm scenario \, (II):} & \quad f \in [0.01, \sqrt{4 \pi}], \quad \mu_{SS} \in [0.1, m_{H_3}] \ {\rm GeV}, \quad  \mu_{1S}= \mu_{2S}  = 10^{-3} \ {\rm GeV},  \\
  {\rm scenario \, (III):} & \quad f \in [0.01, 0.8], \quad \mu_{2S} \in [0.1, m_{H_3}] \ {\rm GeV}, \quad  \mu_{1S}= \mu_{SS}  = 10^{-3} \ {\rm GeV},  \\
 {\rm scenario \, (IV):} & \quad f \in [0.01, 0.8], \quad \mu_{1S} \in [0.1, m_{H_3}] \ {\rm GeV}, \quad  \mu_{2S}= \mu_{SS}  = 10^{-3} \ {\rm GeV}, \nonumber \\
 & \quad m_{H_1} = m_{H^\pm} = m_A \in [70, m_\psi] \ {\rm GeV},
\end{align}
where we assumed $m_{H_1} =m_{H^\pm} =m_A$ for simplicity and they are taken to be larger than $m_\psi$ in scenario (I) to (III).

%%%%%%%%%%%%%%%%%%%%%%%%%%%%%%%%%%%%%%%%%%%%%%%%%%%%%%%%%
\begin{figure}[t] 
%\begin{center}
\includegraphics[width=70mm]{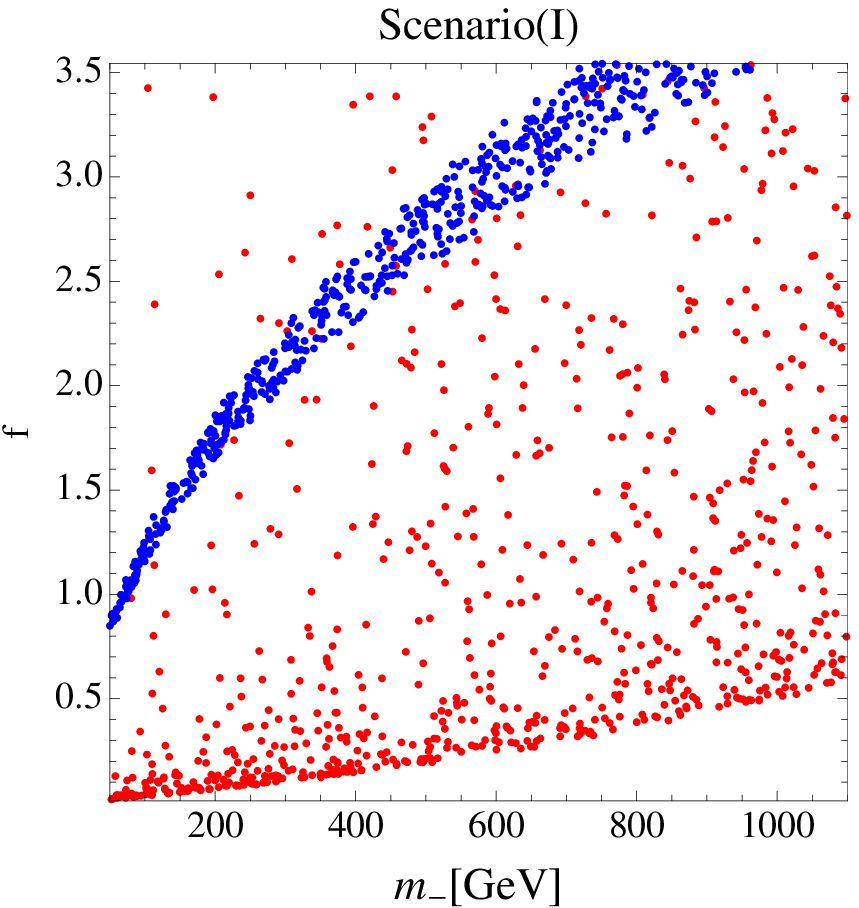} 
\includegraphics[width=70mm]{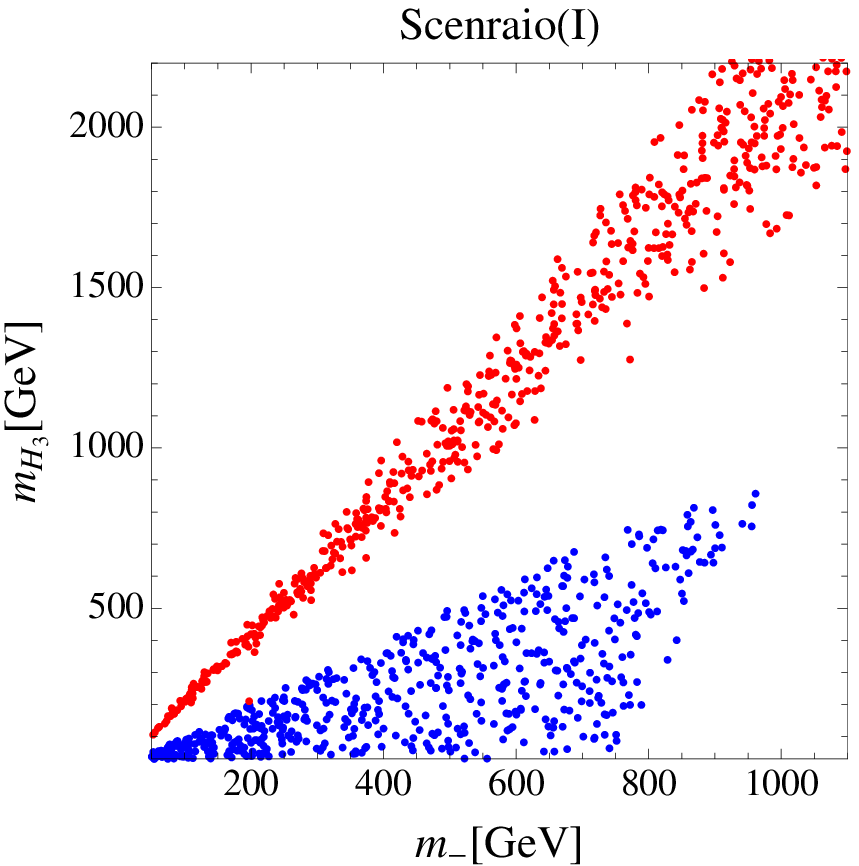} 
%\end{center}
% 
\caption{ The parameter points providing required relic density of DM in scenario (I) where red and blue points
  correspond to the case of $ m_- + m_+ > m_{H_3} > m_-$ and $m_- > m_{H_3}$ respectively. 
%  {\color{red} COULD YOU CHANGE THE  X-AXIS LABEL TO $m_-$ (ALSO IN FIGS. 3.4.5)?}
\label{fig:relic1}}
\end{figure}
%%%%%%%%%%%%%%%%%%%%%%%%%%%%%%%%%%%%%%%%%%%%%%%%%%%%%%%%%%
In Fig.~\ref{fig:relic1}, we show parameter points which explain the observed relic density of DM for scenario (I) where red and blue points correspond to the case of 
(a) $ m_{-} + m_+ > m_{H_3} > m_-$ and (b)$m_- > m_{H_3}$.
We find that the case of {$m_{H_3} >  m_- + m_+$} cannot provide observed relic density with $f < \sqrt{4 \pi}$ since only $\psi_\pm \psi_\pm \to a a$ channel is allowed. 
%{\bf The case (a) allows wide parameter space while case (b) provide narrower allowed region in the $(m_-,f)$-plane since depending on  $f$  various
 % processes contribute  in the case (a) as will be explained below while only $t$-channel 
%$\psi_- \psi_- \to H_3 H_3$ in Fig.~\ref{fig:DM}-(A) is dominant in the case (b). }
In the case (a),  $\psi_- \psi_- \to H_3 \to a a$ in Fig.~\ref{fig:DM}-(B) dominates and the region near resonance, $m_{H_3} \sim 2 m_-$, is
preferred as shown in the right plot of Fig.~\ref{fig:relic1} where $| m_{H_3}- 2m_-|/m_{H_3} \lesssim 10$-$30 \%$ is required. 
{Also in the case (a), for smaller $f$, $t$-channel coannihilation process $\psi_- \psi_+ \to H_3 a$ is
  enhanced near threshold $m_{H_3} \simeq m_{-}+m_{+}$ due to the $t$-channel propagator of $\psi_\pm$ contributes
  $1/(m_+^2 -m_-^2)$ factor to the amplitude. 
%while allowed region for larger $f$ require resonant enhancement of $s$-channel annihilation process $\psi_- \psi_- \to H_3 \to aa $.  
%We also note that for small $f$ coannihilation process is effective due to small mass difference between $\psi_-$ and $\psi_+$.
}
In the case (b), the relevant process is coannihilation $\psi_- \psi_+ \to H_3 a$ as well as $\psi_- \psi_-
  \to H_3 H_3, a a$, shown in Fig.~\ref{fig:DM}-(A),(B).
 The case (a) allows wide parameter space than the case (b) in the $(m_-,f)$-plane simply due to
  resonance dominance in the case (a).

We find that the allowed parameter points for scenario (II) is similar to scenario (I) since new contribution from the
process $\psi_- \psi_- \to H_3 \to H_3 H_3$ is subdominant. 
The allowed region in $(m_-,f)$-plane becomes slightly wider due to new contribution for $m_- > m_{H_3}$ while most of $\mu_{SS}$ region can be allowed. Since the result is similar to that of scenario (I) we omit the plot for scenario (II).

\begin{figure}[t] 
%\begin{center}
\includegraphics[width=70mm]{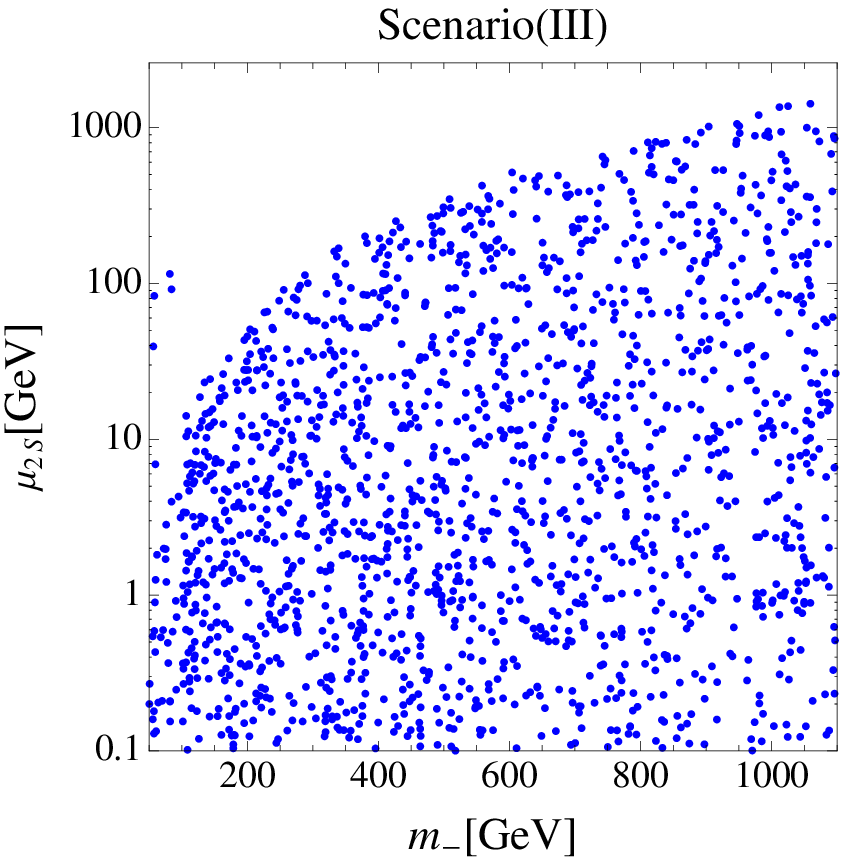} 
\includegraphics[width=70mm]{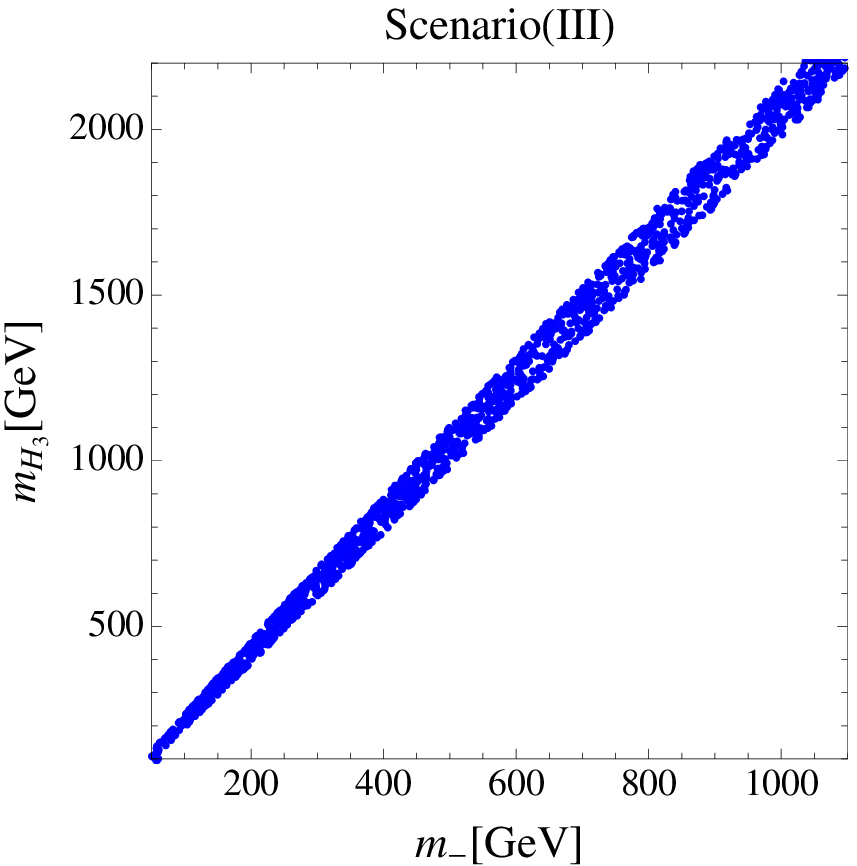} 
\includegraphics[width=70mm]{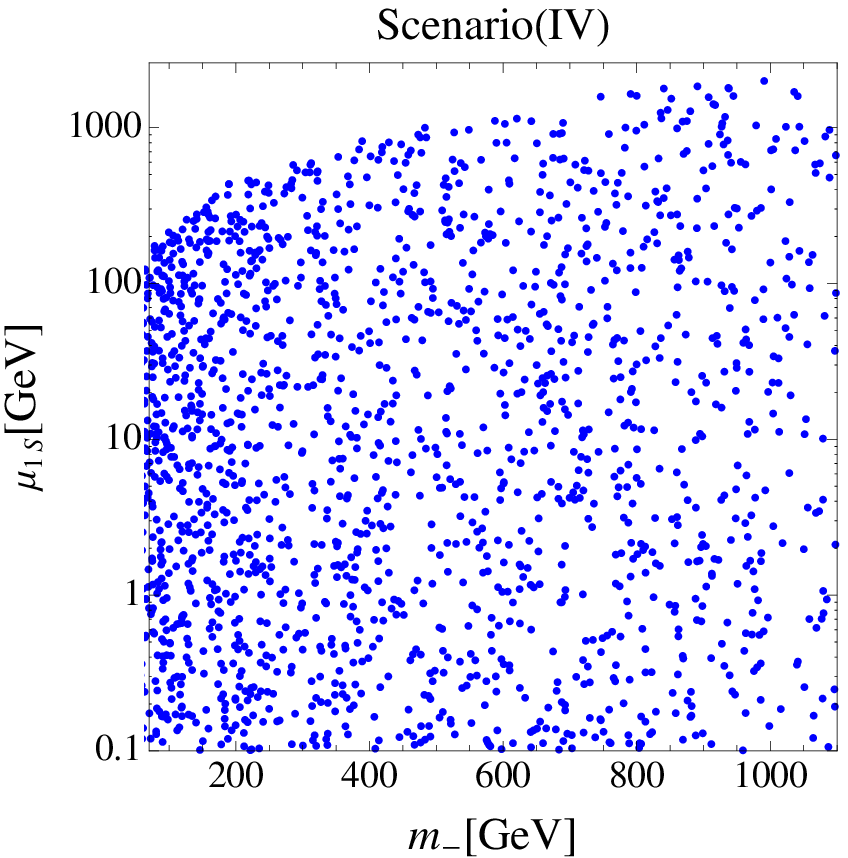} 
\includegraphics[width=70mm]{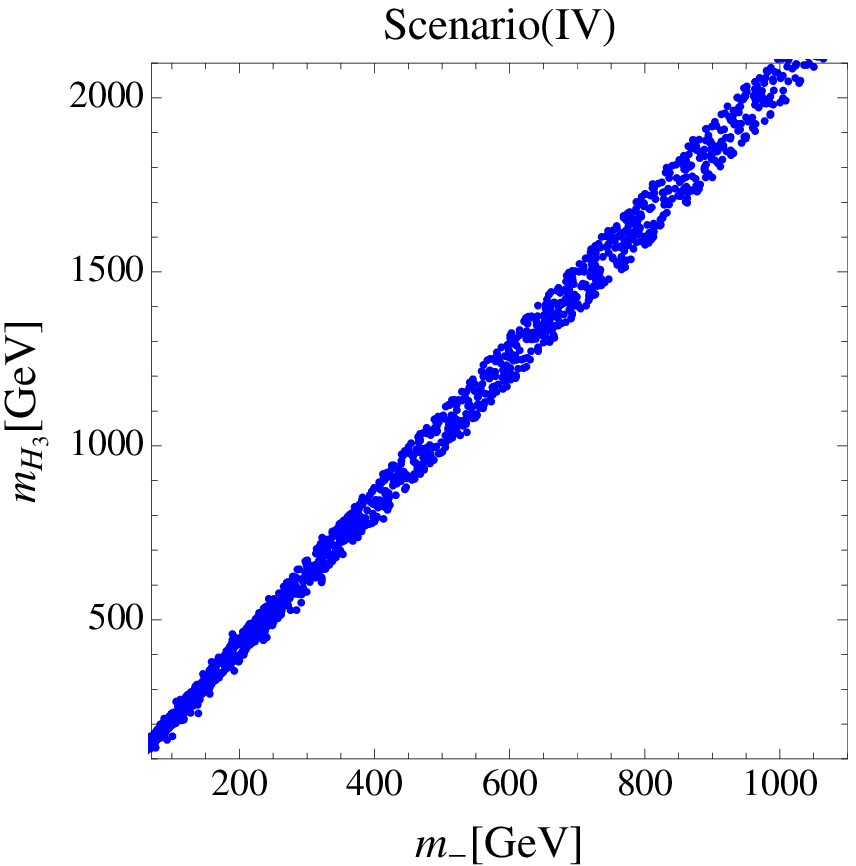} 
%\end{center}
% 
\caption{ The parameter points providing required relic density of DM in scenarios (II) and (III) on the $m_\psi$-$\mu_{SS(2S)}$ plane (left plot) and the $m_\psi$-$m_{H_3}$ plane (right plot). 
\label{fig:relic2}}
\end{figure}

The allowed parameter points for scenario (III) and (IV) are given in Fig.~\ref{fig:relic2} in $(m_-,\mu_{2S(1S)})$- and $(m_-,m_{H_3})$-plane.
We find that parameter space with $m_{H_3} \sim 2 m_-$  as can be seen from Fig.~\ref{fig:DM}-(C) and (D)  can
explain the relic density since resonant enhancement is required to achieve sufficient annihilation cross section where $| m_{H_3}- 2m_-|/m_{H_3} \lesssim 10 \%$ is required.
For the resonant region, wide range of $\mu_{2S(1S)}$ is allowed as shown in left plots of Fig.~\ref{fig:relic2}.
For scenario (III), parameter space with large value of $\mu_{2S}$ is constrained by constraint from mixing angle $ \sin
\theta < 0.2$ and invisible decay branching ratio of SM Higgs. In addition, larger resonant enhancement is required to
obtain sufficient annihilation cross section.
In scenario (IV),  
%the parameter dependence is almost same as scenario (II) replacing $\mu_{SS}$ to $\mu_{1S}$ and we omit corresponding plot.
also dependence on the value of $m_{H_1}$ is small unless it is not very close to that of $m_-$.

%%%%%%%%%%%%%%%%%%%%%%%%%%%%%%%%%%%%%%%%%%%%%%%%%%%%%%%%%
%%%%%%%%%%%%%%%%%%%%%%%%%%%%%%%%%%%%%%%%%%%%%%%%%%%%%%%%%%

\subsection{Direct detection}
\begin{figure}[t] 
\begin{center}
\includegraphics[width=70mm]{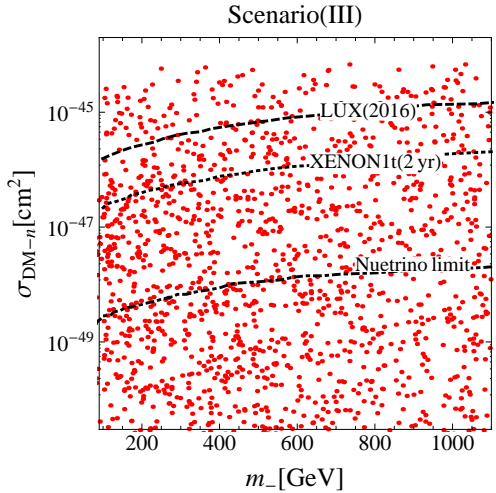} 
\end{center}
 
\caption{ The DM-nucleon scattering cross section for scenario (III) which is compared with current constraint by LUX~\cite{Akerib:2016vxi} and future prospect by XENON 1t~\cite{Aprile:2015uzo}. 
\label{fig:DD}}
\end{figure}

Here we discuss direct detection of DM in our model focusing on our scenario (III) since $\rho$-$h_2$ mixing is negligibly small in other scenarios.
The DM-nucleon scattering is induced by the SM Higgs exchanging process via mixing effect in scalar sector in our model, which is calculated in non-relativistic limit. 
We obtain the following effective Lagrangian by integrating out $h$ and $H_3$;
%\begin{equation}
%{\cal L_{\rm eff}} = \sum_{q} \frac{f m_q s_\theta c_\theta}{2 \sqrt{2} v} \left( \frac{1}{m_{h}^2} - \frac{1}{m_{H_3}^2} \right) (\bar \psi^c \psi + \bar \psi \psi^c) \bar q  q
%\end{equation}
\begin{equation}
{\cal L_{\rm eff}} = \sum_{q} \frac{f m_q s_\theta c_\theta}{2 \sqrt{2} v} \left( \frac{1}{m_{h}^2} - \frac{1}{m_{H_3}^2} \right) \bar \psi_- \psi_-\bar q  q
\end{equation}
where $s_\theta(c_\theta) = \sin \theta (\cos \theta)$, $q$ and $m_q$ denote the corresponding quark field and the quark mass respectively, and the sum is over all quark flavors.
The effective Lagrangian can be rewritten as $\psi$-nucleon (N) interaction:
%\begin{equation}
%{\cal L_{\rm eff}} =  \frac{ f_N m_N f s_\theta c_\theta}{2 \sqrt{2} v} \left( \frac{1}{m_{h}^2} - \frac{1}{m_{H_3}^2} \right) (\bar \psi^c \psi + \bar \psi \psi^c) \bar N  N,
%\end{equation}
\begin{equation}
{\cal L_{\rm eff}} =  \frac{ f_N m_N f s_\theta c_\theta}{2 \sqrt{2} v} \left( \frac{1}{m_{h}^2} - \frac{1}{m_{H_3}^2} \right) \bar \psi_- \psi_-\bar N  N,
\end{equation}
where the effective coupling constant $f_N$ is obtained by
\begin{equation}
f_N = \sum_q f_q^N = \sum_q \frac{m_q}{m_N} \langle N | \bar q q | N \rangle.
\end{equation}
Here we replace the heavy quark contribution by the gluon contributions such that~\cite{Baek:2014jga}
\begin{align}
\sum_{q=c,b,t}  f_q^N = {1 \over m_N} \sum_{q=c,b,t} \langle N | \left(-{ \alpha_s\over 12 \pi} G^a_{\mu\nu}
  G^{a\mu\nu}\right)| N \rangle,
\label{eq:f_Q}
\end{align}
which is obtained by calculating the triangle diagram.
The trace of the stress energy tensor is written as follows by considering the scale anomaly;
\begin{align}
\theta^\mu_\mu =m_N \bar{N} N = \sum_q m_q \bar{q} q - {7 \alpha_s \over 8 \pi} G^a_{\mu\nu} G^{a\mu\nu}.
\label{eq:stressE}
\end{align}
Combining Eqs.~(\ref{eq:f_Q}) and (\ref{eq:stressE}), we obtain 
\begin{align}
\sum_{q=c,b,t} f_q^N = \frac{2}{9} \left( 1 - \sum_{q = u,d,s} f_q^N \right),
\end{align}
which provides
\begin{align}
f_N = \frac29+\frac{7}{9}\sum_{q=u,d,s}f_{q}^N.
\end{align}
Finally the spin independent $\psi$-$N$ scattering cross section is given by~\cite{Hisano:2015bma}
%\begin{equation}
%\sigma_{\rm SI}(\psi N \to \psi N) = \frac{1}{8\pi} \frac{\mu_{N\psi}^2 f_N^2 m_N^2 f^2 s_\theta^2 c_\theta^2}{v^2} \left( \frac{1}{m_{h}^2} - \frac{1}{m_{H_3}^2} \right)^2
%\end{equation}
\begin{equation}
\sigma_{\rm SI}(\psi N \to \psi N) = \frac{1}{2\pi} \frac{\mu_{N\psi}^2 f_N^2 m_N^2 f^2 s_\theta^2 c_\theta^2}{v^2} \left( \frac{1}{m_{h}^2} - \frac{1}{m_{H_3}^2} \right)^2
\end{equation}
where $m_N$ is the nucleon mass and $\mu_{N \psi} = m_N m_-/(m_N + m_-)$ is the reduced mass of nucleon and DM.
For simplicity, we estimate DM-neutron scattering cross section since that of DM-proton is almost the same.
In this case, we apply $f_n \simeq 0.287$(with $f_u^n = 0.0110$, $f_d^n = 0.0273$, $f_s^b = 0.0447$) for the sum of the contributions of partons to the mass fraction of neutron~\cite{Belanger:2013oya}.
The Fig.~\ref{fig:DD} shows the DM-nucleon scattering cross section for the allowed parameter sets in scenario (III); for other scenarios the cross section is negligibly small due to small mixing angle $\theta$. We find that the cross section is mostly smaller than current constraint from LUX~\cite{Akerib:2016vxi} (few parameter space is excluded), and some parameter sets would be tested in future direct detection experiments~\cite{Aprile:2015uzo}.

%%%%%%%%%%%%%%%%%%%%%%%%%%%%%%%%%%%%%%%%%%%%%%%%%%%%%%%%%
%%%%%%%%%%%%%%%%%%%%%%%%%%%%%%%%%%%%%%%%%%%%%%%%%%%%%%%%%%

\subsection{Indirect detection}

%%%%%%%%%%%%%%%%%%%%%%%%%%%%%%%%%%%%%%%%%%%%%%%%%%%%%%%%%
\begin{figure}[t] 
\begin{center}
\includegraphics[width=70mm]{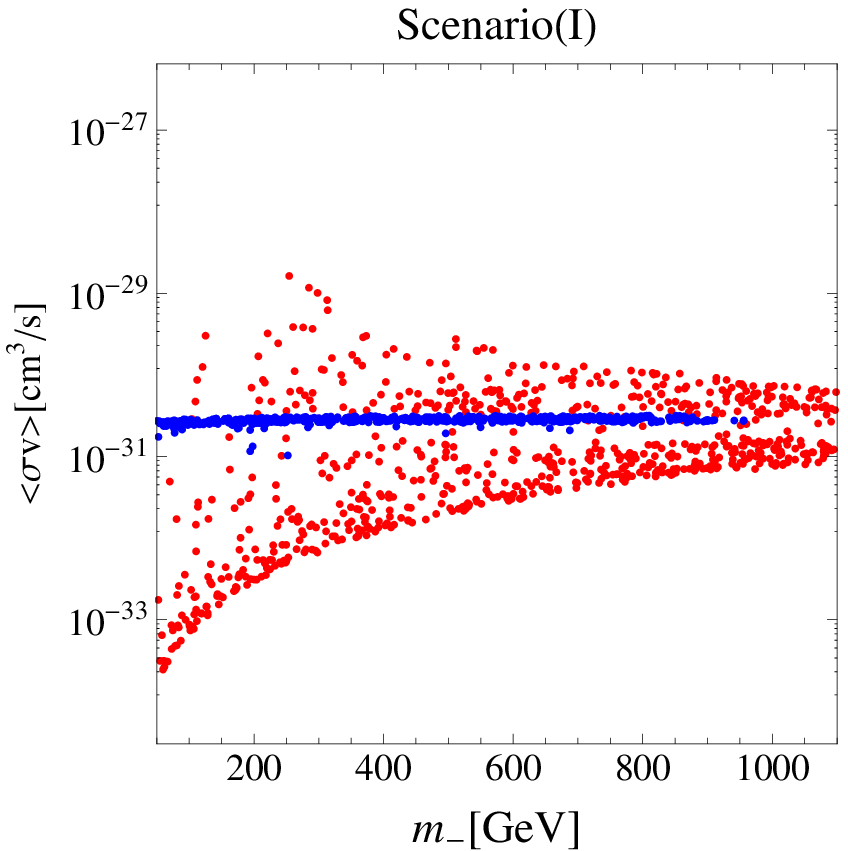} 
\includegraphics[width=70mm]{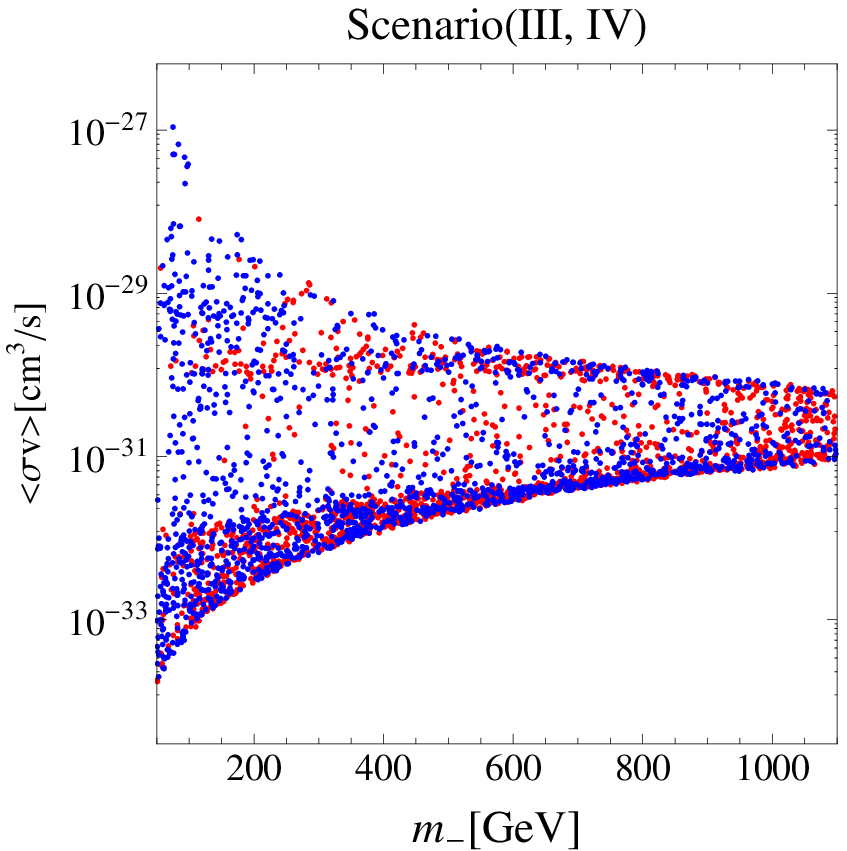} 
\end{center}
 
\caption{ The thermally averaged DM annihilation cross section at current Universe for parameter sets which provides observed relic density.
In the left plot, colors of points correspond to that in Fig.~\ref{fig:relic1}. In the right plot, red and blue points correspond to scenario (III) and (IV) respectively.
\label{fig:ID}}
\end{figure}
%%%%%%%%%%%%%%%%%%%%%%%%%%%%%%%%%%%%%%%%%%%%%%%%%%%%%%%%%%

Here we discuss possibility of indirect detection in our model. 
The thermally averaged cross section in current Universe is estimated with {\tt micrOMEGAs 4.3.1 } applying allowed parameter sets.
The Fig.~\ref{fig:ID} shows the cross section for scenario (I) and scenarios (III,IV) in left and right panel respectively; the scenario (II) provide same feature as scenario (I) and the corresponding plot is omitted here. 

For scenario (I), colors of points correspond to that of in Fig.~\ref{fig:relic1}. We find that the cross section is suppressed since the amplitude of the process decreases as momentum of DM decreases. %while the cross section for $\psi_- \psi_- \to H_3 H_3 (H_3 a)$ is around the value required by relic density.
The cross section for $\psi_- \psi_- \to H_3 H_3$ does not change much while that for $\psi_- \psi_- \to H_3 a (aa)$ has wide range of value since resonant region $m_{H_3} \sim 2 m_-$ is required in the latter case and the current cross section can be much different from that in freeze out era; the case of $m_{H_3} \simeq (\lesssim) 2 m_-$ induce large Breit-Wigner enhancement while the case of $m_{H_3} \gtrsim 2 m_-$ does not induce large enhancement and the cross section is suppressed as the amplitude decreases as DM momentum.
The $H_3$ further decays into $hh$ and SM particles via the effect of mixing with SM Higgs which lead $\gamma$-ray spectrum. 
%Then some parameter space with $\langle \sigma v \rangle \sim 10^{-26} {\rm cm}^3/s$ could be tested by $\gamma$-ray search experiments such as Fermi-LAT~\cite{Ackermann:2015zua}. 
Since the cross section is small, $\gamma$-ray flux is free from current constraint and 
the $\gamma$-ray spectrum depends on decay pattern of $H_3$ and detailed analysis is beyond the scope of this paper.
The scenario (II) provide same result as scenario (I) since annihilation processes are almost same.

For scenario (III) and (IV), the $s$-channel processes with $\mu_{2S}$ and $\mu_{1S}$ can be also enhanced.  
The process $\psi_- \psi_- \to H_3 \to \{h h \}$ is $\lesssim 10^{-28} {\rm cm}^2/s$ due to constraint on $\mu_{2S}$ from mixing with $H_3$ and SM Higgs.
Note that due to resonant enhancement the cross section can be  $\sim 10^{-27} {\rm cm}^2/s$ for the processe $\psi_- \psi_- \to H_3 \to \{H_1 H_1, AA, H^+ H^- \}$ with $m_{-} \lesssim 150$ GeV in scenario (IV) which can be tested by $\gamma$-ray search experiments such as Fermi-LAT~\cite{Ackermann:2015zua} since $H^\pm$ decay into charged leptons. 
%It is also worth to mention that this case contribute to the excess of the gamma-ray spectrum from the Galactic Center by the process $\psi_- \psi_- \to H_3 H_3 \to b b \bar b \bar b$ when DM mass is around 40-80 GeV~\cite{Ko:2014gha,Ko:2014loa, Chen:2015nea}.
%Thus Fermi-LAT data can also constrain some parameter space which provides the enhancement effect. 
The decays of $\{H_1, A, H^\pm \}$ also provide neutrino flux, which is much smaller than current constraint by High energy neutrino search such as IceCube~\cite{Abbasi:2011eq,Aartsen:2013mla}, and It would be tested in future observation.
%For scenario (III), the current cross section is mostly suppressed since $m_{H_3} \gtrsim 2 m_\psi$ is preferred by the relic density.

%%%%%%%%%%%%%%%%%%%%%%%%%%%%%%%%%%%%%%%%%%%%%%%%%%%%%%%%%%%%%%%%%%%%%%%%%%%%
\section{ Conclusions and discussions}
\label{sec:con}
We have studied a dark matter model in which neutrinos get Dirac masses. The global $U(1)_X$ symmetry forbids the
Majorana mass terms of the right-handed neutrinos, thereby allowing the Dirac masses for the neutrinos.
The same symmetry, broken down to a discrete $Z_2$ symmetry, guarantees the stability of a dark matter candidate which
is a hidden sector fermion charged under the global $U(1)_X$. The spontaneous symmetry breaking of $U(1)_X$ occurs due
to VEV, $v_S$, of a hidden sector scalar $S$ whose pseudo-scalar component becomes Goldstone boson, providing a new channel to
the DM annihilations.

We considered four scenarios depending on the size of coupling constants, $f$, $\lambda_{SS} v_S$, $\lambda_{2S} v_S $,
and $\lambda_{1S} v_S$ which regulate the interaction strength of DM and $S$, self-coupling of $S$, SM Higgs and $S$, and scalar doublet for
neutrinos and $S$, respectively. In scenario (I), we assumed $f$ can be large while suppressing $\lambda_{SS} v_S$,
$\lambda_{2S} v_S$, and $\lambda_{1S} v_S$.
In scenarios (II), (III), (IV), we suppressed $f< 0.8$, allowing large $\lambda_{SS} v_S$, $\lambda_{2S} v_S$, and
$\lambda_{1S} v_S$, respectively.

In scenarios (I) and (II), depending on the DM mass, coupling $f \gtrsim 0.05$ can explain the current DM relic abundance.
In scenarios (III) and (IV), the DM relic density can be accommodated near the resonance, $2 m_- \approx m_{H_3}$, where the
DM annihilation cross section is enhanced.

Only scenario (III) has tree-level contribution to the direct detection via dark-scalar mixing with the SM Higgs boson.
Even in this case the direct detection cross section is marginal or well below the current LUX bound due to small mixing as observed
at the LHC.

We also investigated the implications of our model on the indirect detection of DM. In scenarios (I) and (II), the
channels, $\psi_- \psi_- \to \{a H_3, H_3 H_3 \}$, are suppressed because the amplitude is momentum-dependent while the channel $\psi_- \psi_- \to a a$ can be sizable due to Breit-Wigner enhancement. However, $aa$ channel can not be detected by the observation.
In scenario (III), the cross section for $hh$ channel is suppressed due to constraint from $H_3$ and SM mixing.
On the other hand, In scenario (IV), with resonant enhancement the
annihilation cross section for $hh$ and $\{H_1H_1, H^+H^-, AA\}$ can be, $\langle \sigma v \rangle \gtrsim 10^{-27} \, {\rm cm^3/s}$, for $m_{\psi} \lesssim 150$ GeV which is in the ballpark of the
sensitivity of experiments such as Fermi-LAT when scalar bosons decay into charged fermions.
%Our model also has all the ingredients for the excess of gamma-ray spectrum from the Galactic Center, such as $\psi_-
%\psi_- \to H_3 H_3 \to b b \bar b\bar b$ for $40 \lesssim m_- \lesssim 80$ GeV.

\section*{Acknowledgments}
This work is supported in part by National Research Foundation of Korea (NRF) Research Grant NRF-2015R1A2A1A05001869 (SB).
\vspace{0.5cm}
%%%%%%%%%%%%%%%%%%%%%%%%%%%%%%%%%%%
%%%%%%%%%%%%%%%%%%%%%%%%%%%%%%%%%%%


\begin{thebibliography}{99}

%\cite{Davidson:2009ha}
\bibitem{Davidson:2009ha} 
  S.~M.~Davidson and H.~E.~Logan,
  %``Dirac neutrinos from a second Higgs doublet,''
  Phys.\ Rev.\ D {\bf 80}, 095008 (2009)
%  doi:10.1103/PhysRevD.80.095008
  [arXiv:0906.3335 [hep-ph]].
  %%CITATION = doi:10.1103/PhysRevD.80.095008;%%
  %59 citations counted in INSPIRE as of 15 Nov 2016
  
  %\cite{Wang:2006jy}
\bibitem{Wang:2006jy} 
  F.~Wang, W.~Wang and J.~M.~Yang,
  %``Split two-Higgs-doublet model and neutrino condensation,''
  Europhys.\ Lett.\  {\bf 76}, 388 (2006)
%  doi:10.1209/epl/i2006-10293-3
  [hep-ph/0601018].
  %%CITATION = doi:10.1209/epl/i2006-10293-3;%%
  %38 citations counted in INSPIRE as of 07 Dec 2016
  
  %\cite{Mambrini:2015sia}
\bibitem{Mambrini:2015sia} 
  Y.~Mambrini, S.~Profumo and F.~S.~Queiroz,
  %``Dark Matter and Global Symmetries,''
  Phys.\ Lett.\ B {\bf 760}, 807 (2016)
%  doi:10.1016/j.physletb.2016.07.076
  [arXiv:1508.06635 [hep-ph]].
  %%CITATION = doi:10.1016/j.physletb.2016.07.076;%%
  %30 citations counted in INSPIRE as of 07 Dec 2016
  
    %\cite{Weinberg:2013kea}
\bibitem{Weinberg:2013kea} 
  S.~Weinberg,
  %``Goldstone Bosons as Fractional Cosmic Neutrinos,''
  Phys.\ Rev.\ Lett.\  {\bf 110}, no. 24, 241301 (2013)
 % doi:10.1103/PhysRevLett.110.241301
  [arXiv:1305.1971 [astro-ph.CO]].
  %%CITATION = doi:10.1103/PhysRevLett.110.241301;%%
  %90 citations counted in INSPIRE as of 21 Nov 2016

\bibitem{tHooft}
G.~'t~ Hooft, ``Naturalness, chiral symmetry, and spontaneous chiral symmetry breaking'', Proceedings of the 1979
Carg\`{e}se Institute on Recent Developments in Gauge Theories, G.~'t~Hooft, {\it et. al.} eds., Plenum Press, New York,
U.S.A (1980).

\bibitem{Baek:2014sda}
  S.~Baek,
  %``3.5 keV X-ray line signal from dark matter decay in local $U(1)_{B-L}$ extension of Zee-Babu model,''
  JHEP {\bf 1508} (2015) 023
%  doi:10.1007/JHEP08(2015)023
  [arXiv:1410.1992 [hep-ph]].
  %%CITATION = doi:10.1007/JHEP08(2015)023;%%

\bibitem{Frampton:2002rn}
  P.~H.~Frampton, M.~C.~Oh and T.~Yoshikawa,
  %``Majorana mass zeroes from triplet VEV without majoron problem,''
  Phys.\ Rev.\ D {\bf 66} (2002) 033007
 % doi:10.1103/PhysRevD.66.033007
  [hep-ph/0204273].
  %%CITATION = doi:10.1103/PhysRevD.66.033007;%%

\bibitem{Chang:1988aa}
  D.~Chang, W.~Y.~Keung and P.~B.~Pal,
  %``Spontaneous Lepton Number Breaking At Electroweak Scale,''
  Phys.\ Rev.\ Lett.\  {\bf 61} (1988) 2420.
%  doi:10.1103/PhysRevLett.61.2420
  %%CITATION = doi:10.1103/PhysRevLett.61.2420;%%



\bibitem{Riess:2016jrr}
  A.~G.~Riess {\it et al.},
  %``A 2.4% Determination of the Local Value of the Hubble Constant,''
  Astrophys.\ J.\  {\bf 826} (2016) no.1,  56
%  doi:10.3847/0004-637X/826/1/56
  [arXiv:1604.01424 [astro-ph.CO]].
  %%CITATION = doi:10.3847/0004-637X/826/1/56;%%

\bibitem{Ko:2016uft}
  P.~Ko and Y.~Tang,
  %``Light dark photon and fermionic dark radiation for the Hubble constant and the structure formation,''
  Phys.\ Lett.\ B {\bf 762} (2016) 462
%  doi:10.1016/j.physletb.2016.10.001
  [arXiv:1608.01083 [hep-ph]].
  %%CITATION = doi:10.1016/j.physletb.2016.10.001;%%



\bibitem{Ade:2015xua}
  P.~A.~R.~Ade {\it et al.} [Planck Collaboration],
  %``Planck 2015 results. XIII. Cosmological parameters,''
  Astron.\ Astrophys.\  {\bf 594} (2016) A13
%  doi:10.1051/0004-6361/201525830
  [arXiv:1502.01589 [astro-ph.CO]].
  %%CITATION = doi:10.1051/0004-6361/201525830;%%
  

  %%%%%%%%%%%%%%%%% Higgs mixing %%%%%%%%%%%%%%%%%%%%%%%%%%%%%%
  
    \bibitem{hdecay}  The numerical analyses on the Higgs decays are performed using the program {\tt HDECAY}: A.~Djouadi, J.~Kalinowski and M.~Spira, Comput. Phys.
Commun. 108 (1998) 56; A. Djouadi, M. Muhlleitner and M. Spira, Acta. Phys. Polon. 
B38 (2007) 635. 
  
  

  
%\cite{Chpoi:2013wga}
\bibitem{Chpoi:2013wga} 
  S.~Choi, S.~Jung and P.~Ko,
  %``Implications of LHC data on 125 GeV Higgs-like boson for the Standard Model and its various extensions,''
  JHEP {\bf 1310}, 225 (2013)
 % doi:10.1007/JHEP10(2013)225
  [arXiv:1307.3948 [hep-ph]].
  %%CITATION = doi:10.1007/JHEP10(2013)225;%%
 
%\cite{Cheung:2015dta}
\bibitem{Cheung:2015dta} 
  K.~Cheung, P.~Ko, J.~S.~Lee and P.~Y.~Tseng,
  %``Bounds on Higgs-Portal models from the LHC Higgs data,''
  JHEP {\bf 1510}, 057 (2015)
%  doi:10.1007/JHEP10(2015)057
  [arXiv:1507.06158 [hep-ph]].
  %%CITATION = doi:10.1007/JHEP10(2015)057;%%

  
  %\cite{Cheung:2015cug}
\bibitem{Cheung:2015cug} 
  K.~Cheung, P.~Ko, J.~S.~Lee, J.~Park and P.~Y.~Tseng,
  %``Higgs precision study of the 750 GeV diphoton resonance and the 125 GeV standard model Higgs boson with Higgs-singlet mixing,''
  Phys.\ Rev.\ D {\bf 94}, no. 3, 033010 (2016)
%  doi:10.1103/PhysRevD.94.033010
  [arXiv:1512.07853 [hep-ph]].
  %%CITATION = doi:10.1103/PhysRevD.94.033010;%%
  %101 citations counted in INSPIRE as of 19 Nov 2016
 
     %\cite{Benbrik:2015evd}
\bibitem{Benbrik:2015evd} 
  R.~Benbrik, C.~H.~Chen and T.~Nomura,
  %``$h,Z\to \ell_i \bar\ell_j$, $\Delta a_{\mu}$, $\tau\to (3\mu,\mu \gamma)$ in generic two-Higgs-doublet models,''
  Phys.\ Rev.\ D {\bf 93}, no. 9, 095004 (2016)
%  doi:10.1103/PhysRevD.93.095004
  [arXiv:1511.08544 [hep-ph]].
  %%CITATION = doi:10.1103/PhysRevD.93.095004;%%
  %11 citations counted in INSPIRE as of 24 Aug 2016

  
  %%%%%%%%%%%%%%%%%%%%%%%%%%%%%%%%%%%%%% limit from invisible Higgs decay %%%%%%%%%
  
  %\cite{Aad:2015txa}
\bibitem{Aad:2015txa} 
  G.~Aad {\it et al.} [ATLAS Collaboration],
  %``Search for invisible decays of a Higgs boson using vector-boson fusion in $pp$ collisions at $\sqrt{s}=8$ TeV with the ATLAS detector,''
  JHEP {\bf 1601}, 172 (2016)
%  doi:10.1007/JHEP01(2016)172
  [arXiv:1508.07869 [hep-ex]].
  %%CITATION = doi:10.1007/JHEP01(2016)172;%%
  %37 citations counted in INSPIRE as of 25 Nov 2016
  
  %\cite{Aad:2015pla}
\bibitem{Aad:2015pla}
  G.~Aad {\it et al.} [ATLAS Collaboration],
  %``Constraints on new phenomena via Higgs boson couplings and invisible decays with the ATLAS detector,''
  JHEP {\bf 1511} (2015) 206
%  doi:10.1007/JHEP11(2015)206
  [arXiv:1509.00672 [hep-ex]].
  %%CITATION = doi:10.1007/JHEP11(2015)206;%%
  %90 citations counted in INSPIRE as of 25 Nov 2016
  
  %\cite{Khachatryan:2016whc}
\bibitem{Khachatryan:2016whc} 
  V.~Khachatryan {\it et al.} [CMS Collaboration],
  %``Searches for invisible decays of the Higgs boson in pp collisions at $\sqrt{s}$ = 7, 8, and 13 TeV,''
  arXiv:1610.09218 [hep-ex].
  %%CITATION = ARXIV:1610.09218;%%
  %2 citations counted in INSPIRE as of 25 Nov 2016
  
  %%%%%%%%%%%%%%%%%%%%%%%%%%%%%%%%%%%%%%%%%%%%%%%%%%%%%%%%
  
  %\cite{Machado:2015sha}
\bibitem{Machado:2015sha} 
  P.~A.~N.~Machado, Y.~F.~Perez, O.~Sumensari, Z.~Tabrizi and R.~Z.~Funchal,
  %``On the Viability of Minimal Neutrinophilic Two-Higgs-Doublet Models,''
  JHEP {\bf 1512}, 160 (2015)
%  doi:10.1007/JHEP12(2015)160
  [arXiv:1507.07550 [hep-ph]].
  %%CITATION = doi:10.1007/JHEP12(2015)160;%%
  %7 citations counted in INSPIRE as of 08 Feb 2017

%\cite{Bertuzzo:2015ada}
\bibitem{Bertuzzo:2015ada} 
  E.~Bertuzzo, Y.~F.~Perez G., O.~Sumensari and R.~Zukanovich Funchal,
  %``Limits on Neutrinophilic Two-Higgs-Doublet Models from Flavor Physics,''
  JHEP {\bf 1601}, 018 (2016)
%  doi:10.1007/JHEP01(2016)018
  [arXiv:1510.04284 [hep-ph]].
  %%CITATION = doi:10.1007/JHEP01(2016)018;%%
  %4 citations counted in INSPIRE as of 08 Feb 2017

  
  %%%%%%%%%%%%%%%%%%%%%%%%%%%%%%%%%%%%%%%%%%%%%%%%%%%%%%%
\bibitem{Lindner:2011it}
  M.~Lindner, D.~Schmidt and T.~Schwetz,
  %``Dark Matter and neutrino masses from global U(1)$_{B−L}$ symmetry breaking,''
  Phys.\ Lett.\ B {\bf 705} (2011) 324
%  doi:10.1016/j.physletb.2011.10.022
  [arXiv:1105.4626 [hep-ph]].
  %%CITATION = doi:10.1016/j.physletb.2011.10.022;%%

\bibitem{Baek:2012ub}
  S.~Baek, P.~Ko, H.~Okada and E.~Senaha,
  %``Can Zee-Babu model implemented with scalar dark matter explain both Fermi/LAT 130 GeV $\gamma$-ray excess and neutrino physics ?,''
  JHEP {\bf 1409} (2014) 153
%  doi:10.1007/JHEP09(2014)153
  [arXiv:1209.1685 [hep-ph]].
  %%CITATION = doi:10.1007/JHEP09(2014)153;%%

\bibitem{Baek:2013ywa}
  S.~Baek and H.~Okada,
  %``Hidden sector dark matter with global $U(1)_X$-symmetry and Fermi-LAT 130 GeV $\gamma$-ray excess,''
  Phys.\ Lett.\ B {\bf 728} (2014) 630
%  doi:10.1016/j.physletb.2013.12.033
  [arXiv:1311.2380 [hep-ph]].
  %%CITATION = doi:10.1016/j.physletb.2013.12.033;%%

%%%%%%%%%%%%%%%%%%%Higgs portal%%%%%%%%%%%%%%%%%%%%%%%%%%%%%%%%%%%%%%%%
\bibitem{Kim:2008pp}
  Y.~G.~Kim, K.~Y.~Lee and S.~Shin,
  %``Singlet fermionic dark matter,''
  JHEP {\bf 0805} (2008) 100
%  doi:10.1088/1126-6708/2008/05/100
  [arXiv:0803.2932 [hep-ph]].
  %%CITATION = doi:10.1088/1126-6708/2008/05/100;%%

\bibitem{Baek:2011aa}
  S.~Baek, P.~Ko and W.~I.~Park,
  %``Search for the Higgs portal to a singlet fermionic dark matter at the LHC,''
  JHEP {\bf 1202} (2012) 047
%  doi:10.1007/JHEP02(2012)047
  [arXiv:1112.1847 [hep-ph]].
  %%CITATION = doi:10.1007/JHEP02(2012)047;%%

\bibitem{Baek:2013qwa}
  S.~Baek, P.~Ko and W.~I.~Park,
  %``Singlet Portal Extensions of the Standard Seesaw Models to a Dark Sector with Local Dark Symmetry,''
  JHEP {\bf 1307} (2013) 013
%  doi:10.1007/JHEP07(2013)013
  [arXiv:1303.4280 [hep-ph]].
  %%CITATION = doi:10.1007/JHEP07(2013)013;%%

\bibitem{Baek:2012se}
  S.~Baek, P.~Ko, W.~I.~Park and E.~Senaha,
  %``Higgs Portal Vector Dark Matter : Revisited,''
  JHEP {\bf 1305} (2013) 036
%  doi:10.1007/JHEP05(2013)036
  [arXiv:1212.2131 [hep-ph]].
  %%CITATION = doi:10.1007/JHEP05(2013)036;%%

\bibitem{Baek:2014jga}
  S.~Baek, P.~Ko and W.~I.~Park,
  %``Invisible Higgs Decay Width vs. Dark Matter Direct Detection Cross Section in Higgs Portal Dark Matter Models,''
  Phys.\ Rev.\ D {\bf 90} (2014) no.5,  055014
  %doi:10.1103/PhysRevD.90.055014
  [arXiv:1405.3530 [hep-ph]].
  %%CITATION = doi:10.1103/PhysRevD.90.055014;%%
  
    %\cite{Chen:2015nea}
\bibitem{Chen:2015nea} 
  C.~H.~Chen and T.~Nomura,
  %``$SU(2)_X$ vector DM and Galactic Center gamma-ray excess,''
  Phys.\ Lett.\ B {\bf 746}, 351 (2015)
%  doi:10.1016/j.physletb.2015.05.027
  [arXiv:1501.07413 [hep-ph]].
  %%CITATION = doi:10.1016/j.physletb.2015.05.027;%%
  %14 citations counted in INSPIRE as of 19 Nov 2016

\bibitem{Baek:2016kud}
  S.~Baek, T.~Nomura and H.~Okada,
  %``An explanation of one-loop induced h → μτ decay,''
  Phys.\ Lett.\ B {\bf 759} (2016) 91
  %doi:10.1016/j.physletb.2016.05.055
  [arXiv:1604.03738 [hep-ph]].
  %%CITATION = doi:10.1016/j.physletb.2016.05.055;%%
  
  %%%%%%%%%%%%%%%%%%%%%%%%%%%%%%%%%%%%%%%%%%%%%%%%%%%%%%%%%%%%


  
      \bibitem{Belanger:2014vza} 
  G.~Belanger, F.~Boudjema, A.~Pukhov and A.~Semenov,
  %``micrOMEGAs4.1: two dark matter candidates,''
  arXiv:1407.6129 [hep-ph].
  %%CITATION = ARXIV:1407.6129;%%
  %1 citations counted in INSPIRE as of 20 Jan 2015
  
    %\cite{Ade:2013zuv}
\bibitem{Ade:2013zuv} 
  P.~A.~R.~Ade {\it et al.}  [Planck Collaboration],
  %``Planck 2013 results. XVI. Cosmological parameters,''
  Astron.\ Astrophys.\  (2014)
  [arXiv:1303.5076 [astro-ph.CO]].
  %%CITATION = ARXIV:1303.5076;%%
  %2596 citations counted in INSPIRE as of 08 Nov 2014
  
   \bibitem{Hisano:2015bma}
  J.~Hisano, R.~Nagai and N.~Nagata,
  %``Effective Theories for Dark Matter Nucleon Scattering,''
  JHEP {\bf 1505} (2015) 037
 % doi:10.1007/JHEP05(2015)037
  [arXiv:1502.02244 [hep-ph]].
  %%CITATION = doi:10.1007/JHEP05(2015)037;%%
 
 %\cite{Belanger:2013oya}
\bibitem{Belanger:2013oya} 
  G.~Belanger, F.~Boudjema, A.~Pukhov and A.~Semenov,
  %``micrOMEGAs_3: A program for calculating dark matter observables,''
  Comput.\ Phys.\ Commun.\  {\bf 185}, 960 (2014)
  [arXiv:1305.0237 [hep-ph]].
  %%CITATION = ARXIV:1305.0237;%%
  %215 citations counted in INSPIRE as of 14 juin 2015
 
  
    %\cite{Akerib:2016vxi}
\bibitem{Akerib:2016vxi} 
  D.~S.~Akerib {\it et al.},
  %``Results from a search for dark matter in LUX with 332 live days of exposure,''
  arXiv:1608.07648 [astro-ph.CO].
  %%CITATION = ARXIV:1608.07648;%%
  %26 citations counted in INSPIRE as of 08 Oct 2016
  
    %\cite{Aprile:2015uzo}
\bibitem{Aprile:2015uzo} 
  E.~Aprile {\it et al.} [XENON Collaboration],
  %``Physics reach of the XENON1T dark matter experiment,''
  JCAP {\bf 1604}, no. 04, 027 (2016)
%  doi:10.1088/1475-7516/2016/04/027
  [arXiv:1512.07501 [physics.ins-det]].
  %%CITATION = doi:10.1088/1475-7516/2016/04/027;%%
  %41 citations counted in INSPIRE as of 13 Jul 2016
  

  
  
  
 
  %\cite{Ackermann:2015zua}
\bibitem{Ackermann:2015zua} 
  M.~Ackermann {\it et al.} [Fermi-LAT Collaboration],
  %``Searching for Dark Matter Annihilation from Milky Way Dwarf Spheroidal Galaxies with Six Years of Fermi Large Area Telescope Data,''
  Phys.\ Rev.\ Lett.\  {\bf 115}, no. 23, 231301 (2015)
%  doi:10.1103/PhysRevLett.115.231301
  [arXiv:1503.02641 [astro-ph.HE]].
  %%CITATION = doi:10.1103/PhysRevLett.115.231301;%%
  %249 citations counted in INSPIRE as of 07 Jul 2016
  
  %%%%%%%%%%%%%%%%%%%%%%%%%%%%%%%%%%%%%%%%
  

%%%%%%%%%%%%%%%%%%%%%%%%%%%%%%%%%%%%%%%%%%%%%%%%%%%%%

%\cite{Abbasi:2011eq}
\bibitem{Abbasi:2011eq}
  R.~Abbasi {\it et al.}  [IceCube Collaboration],
  %``Search for Dark Matter from the Galactic Halo with the IceCube Neutrino Observatory,''
  Phys.\ Rev.\ D {\bf 84} (2011) 022004
  [arXiv:1101.3349 [astro-ph.HE]].
  %%CITATION = ARXIV:1101.3349;%%
  %64 citations counted in INSPIRE as of 12 Mar 2014
    
%\cite{Aartsen:2013mla}
\bibitem{Aartsen:2013mla} 
  M.~G.~Aartsen {\it et al.}  [IceCube Collaboration],
  %``The IceCube Neutrino Observatory Part IV: Searches for Dark Matter and Exotic Particles,''
  arXiv:1309.7007 [astro-ph.HE].
  %%CITATION = ARXIV:1309.7007;%%
  %6 citations counted in INSPIRE as of 12 Mar 2014

  
\end{thebibliography}
\end{document}